\documentclass[11pt,tightenlines,aps,prd,groupedaddress]{revtex4}

\usepackage{amsmath,amssymb,bm}

\usepackage{graphicx}
\usepackage{amsfonts}
\usepackage{amssymb}
\usepackage{amsbsy}
\usepackage{amsmath}
\usepackage{latexsym}
\usepackage{bm}
\usepackage{color}
\usepackage{comment}

\newcommand*{\di}{\partial}

\def\b{\bar}
\def\l{\left}
\def\r{\right}
\def\p {\partial}

\def\be {\begin{eqnarray}}
\def\ee {\end{eqnarray}}
\def\nn {\nonumber}

\begin{document}

\title{Linearized gravity with matter time}

\author{Masooma Ali}
\author{Viqar Husain}
\author{Shohreh Rahmati}
\author{Jonathan Ziprick}
\email{vhusain@unb.ca}

\affiliation{University of New Brunswick\\
Department of Mathematics and Statistics\\
Fredericton, NB E3B 5A3, Canada}


\begin{abstract}
 
We study general relativity with pressureless dust  in the canonical formulation, with the dust field  chosen as a matter-time gauge. The resulting theory has three physical degrees of freedom in the metric field. The linearized canonical theory reveals two graviton modes and  a scalar mode. We find that the graviton modes remain Lorentz covariant despite the time gauge, and that the scalar mode is ultralocal.  We  discuss also a modification of the theory to  include a parameter in the Hamiltonian analogous to that in Horava-Lifshitz models.  In this case the scalar mode is no longer ultralocal, and acquires a propagation speed dependent on the deformation parameter. 

\end{abstract}

\maketitle

\section{Introduction}

In the Hamiltonian formulation of general relativity there is an understanding of how  graviton modes arise. This  parallels the more familiar covariant analysis of linearized theory. The analysis  starts with the expansion of the canonical Arnowittt-Deser-Misner (ADM)  phase space variables and constraints of general relativity (GR) around the flat spacetime solution,  followed by imposition of  the  canonical transverse traceless (TT) gauge  condition \cite{Arnowitt:1962hi}. The solution of the constraints in this gauge gives  the two (unconstrained) graviton degrees of freedom. (See e.g. \cite{Hanson:1976cn}  for a pedagogical review of the ADM analysis). 

The ADM analysis of the TT gauge is for vacuum  general relativity. If there is matter coupling it is possible to  set up a linearized canonical theory with matter perturbations, as is done for example in the theory of cosmological perturbations \cite{Langlois:1993}. The standard analysis however still uses the TT gauge, which taken together are  four conditions that fix four coordinates. Importantly, these gauge choices  are  conditions only on the gravitational phase space variables; there are no restrictions on the matter sector as far as the fixing of general coordinate invariance is concerned. 

Although the TT gauge is very useful and meshes well with the covariant analyses of gravitational perturbation theory, it is only one choice.  In a Hamiltonian setting with general relativity coupled to matter, it is clearly possible to use matter degrees of freedom in making coordinate gauge choices. This is not  usually done because the interpretation of gravitational waves as spin 2 fields on a background spacetime is lost.  Furthermore, it is apparent that if matter degrees of freedom (``matter reference systems")  are used to fix spacetime diffeomorphism freedom, then the gauge fixed theory has additional local degrees of freedom in the geometry sector, which makes it harder to  interpret physically.     

We nevertheless consider this possibility in the specific setting of general relativity  coupled to pressureless dust, and any other matter field. This is a special case of the matter reference system used by Brown and Kuchar  \cite{Brown:1994py}, who used a more general 4-component  dust field to set all four coordinate conditions. Our analysis uses the dust time gauge, where the dust scalar is set as the time coordinate.   We develop  this  idea  in part because it leads to some novel results which have a bearing on discussions of Lorentz violation and quantum gravity; after a complete gauge  fixing, we find that the scalar field degree of freedom is manifest as a scalar mode in the spatial metric, and the interpretation of gravitational waves as spin 2 fields on the background is preserved. 

  There is recent related work in two directions. These are the Einstein-Aether models \cite{Eling:2004dk}, where a dynamical vector field of timelike norm is added to the GR action. A linearized analysis of these models has been performed, with the result that the graviton modes decouple from the aether modes \cite{Jacobson:2004ts}.  The other model is the  so-called mimetic gravity \cite{Chamseddine:2013kea}, where the conformal mode of the spacetime metric is encoded as a scalar field with an arbitrary potential. This extra mode in the gravitational field represents  self-interacting matter with arbitrary  potential \cite{Lim:2010yk, Chamseddine:2014}, and has been used to model inflationary and bouncing cosmologies. When the potential vanishes, the canonical theory  is equivalent to general relativity  coupled to  pressureless dust, which is studied as a quantum gravity model  in  \cite{Husain:2011tk}.   Given these analogies, it is  potentially useful to consider this  paper in the larger context of Einstein-Aether \cite{Jacobson:2015mra} and mimetic gravity theories.

In the next section we consider the GR+dust  theory and review the use of the dust time gauge in the ADM canonical framework \cite{Husain:2011tk}.  In Section II we present an analysis of the linearized theory about flat spacetime in the Hamiltonian theory. This section contains the main result: of the three physical degrees of freedom in the metric, two are the graviton modes, and the third degree is ultralocal.  We also comment on extensions of the theory to include the Horava-Lifshitz (HL) parameter \cite{Horava:2009uw, Horava:2010zj}  in the physical Hamiltonian; (for a Hamltonian formulation of HL gravity and  review see \cite{Donnelly:2011df} and \cite{Visser:2011mf}). We find that if the HL deformation parameter is different from its GR value,  the two graviton mode equations are unaffected, but the third has  propagation  speed determined by this parameter. We conclude in Section IV with  a summary and discussion of the role of dust time gauge for quantization.

\section{Action and Hamiltonian theory}

We consider GR coupled to pressureless dust and any other arbitrary  matter field,
\be
S=\frac{1}{2\pi}\int{d^{4}x\sqrt{-g}R}-\frac{1}{4\pi}\int{d^{4}x\sqrt{-g}\ m(g^{\mu\nu}\di_{\mu}\phi\di_{\nu}\phi+1)}  + \int d^4x\  {\cal L}_M(\chi).
\ee
The  second term is the dust action, and the last is an arbitrary matter Lagrangian. Variation with respect to $m$ gives the condition that the dust field $\phi$ has timelike gradient.

The ADM canonical theory obtained from this action is 
\be
S=\frac{1}{2\pi}\int{dt \ d^{3}x\left(\tilde{\pi}^{ab}\dot{q}_{ab}+p_{\phi}\dot{\phi} + p_\chi \dot{\chi}-N\mathcal{H}-N^{a}\mathcal{C}_{a}\right)},
\label{can-act}
\ee
where the pairs  $(q_{ab},\tilde{\pi}^{ab})$ and $(\phi, p_{\phi})$ are respectively the phase space variables of gravity and dust.  The matter fields are symbolically denoted by $(\chi, p_{\chi})$, although the number of fields and their tensorial structures will depend upon the choice of the matter Lagrangian. The lapse and shift functions,  $N$ and $N^{a}$ are the coefficients of the Hamiltonian and diffeomorphism constraints
\be
\label{HG}
\mathcal{H} &=&\mathcal{H}^{G}+\mathcal{H}^{D}  + \mathcal{H}^{M},\\
\mathcal{C}_{a}&=&\mathcal{C}^{G}_{a}+\mathcal{C}^{D}_{a} +\mathcal{C}^{M}_{a}\nn\\
 &=&-2D_{b}\tilde{\pi}^{b}_{a}+p_{\phi}\di_{a}\phi + \mathcal{C}^{M}_{a},
\ee
where
\be
\mathcal{H}^{G}&=& -\sqrt{q} R^{(3)}+\frac{1}{\sqrt{q}} \left(\tilde{\pi}^{ab}\tilde{\pi}_{ab}- \frac{1}{2}\tilde{\pi}^{2}\right) ,\\
\mathcal{H}^{D}&=&\frac{1}{2}\left(\frac{p_{\phi}^{2}}{m\sqrt{q}}+m\sqrt{q}(q^{ab}\di_{a}\phi\di_{b}\phi+1)\right).
\ee
The trace of the gravitational momentum is $\tilde{\pi}=q_{ab}\tilde{\pi}^{ab}$, $R^{(3)}$ is the scalar curvature of the spatial hypersurfaces, $D_a$ is the metric compatible covariant derivative associated with $q_{ab}$, and $\mathcal{H}^{M}$ is the (non-dust) matter Hamiltonian.   

The momentum conjugate to the field $m$ is zero since it appears as a Lagrange multiplier in the covariant action. At this point one could enlarge the phase space to treat $m$ and its conjugate momentum as independent degrees of freedom, subsequently eliminating them by gauge fixing. However, it is more straightforward to vary the term $\mathcal{H}^{D}$ in the canonical action with respect to $m$, and use the resulting equation of motion: 
\be
\label{m}
m=\pm\frac{p_{\phi}}{\sqrt{q(q^{ab}\di_{a}\phi\di_{b}\phi+1)}}.
\ee
This can then be substituted back into $\mathcal{H}^{D}$ to give
\be
\mathcal{H}^{D}=  \pm\  p_\phi  \sqrt{q^{ab}\di_{a}\phi\di_{b}\phi+1},
\ee
leaving a canonical action for  $(q_{ab},\tilde{\pi}^{ab})$, $(\phi, p_{\phi})$ and the (non-dust) matter phase space variables.
It is readily verified that the constraints remain  first class with this elimination of $m$. We will see in the gauge fixing below how the sign is selected.

\subsection{Dust time gauge}

We now partially reduce the theory  by fixing a time gauge and solving the Hamiltonian constraint to obtain a physical Hamiltonian. We use the dust time gauge  \cite{Husain:2011tk,Swiezewski:2013jza} which equates the physical time with level values of the scalar field, 
\be
\label{gauge}
\lambda\equiv \phi-t\approx0.
\ee
This has a nonzero Poisson bracket with the Hamiltonian constraint, so this pair of constraints is second class. Requiring that the gauge condition be preserved in time gives an equation for the lapse function:
\be
&&1 = \dot{\phi}= \left. \left\{\phi, \int d^3x \left (N  \mathcal{H}  + N^a \mathcal{C}_{a}\right) \right\} \right|_{\phi=t}  =   N\  \frac{p_\phi}{m\sqrt{q}}\nn\\
 && \implies N= \frac{m\sqrt{q}}{p_\phi} \ .
\ee
Substituting     $\phi=t$  into (\ref{m}) gives $ p_\phi=\pm m\sqrt{q}$, which by the last relation  leads to  $N=\pm 1$.  The sign of the lapse function determines whether the evolution is forward ($N=+1$) or backward ($N=-1$) in time. We select the positive sign which fixes the above ambiguity in the Hamiltonian constraint, yielding $\mathcal{H}^{D}= +p_\phi$.

With these results, solving the Hamiltonian constraint  gives 
\be
p_\phi = - \left( \mathcal{H}^{G} + \mathcal{H}^{M} \right).
\ee
Substituting this and the gauge condition (\ref{gauge}) into (\ref{can-act}) gives  the gauge fixed action
\be
S_{GF}=\frac{1}{2\pi}\int{ dt \ d^3x  \left[\tilde{\pi}^{ab}\dot{q}_{ab} +p_\chi\dot{\chi}  -( \mathcal{H}^{G} + \mathcal{H}^{M}) -N^{a}(\mathcal{C}^{G}_{a} +\mathcal{C}^{M}_{a})  \right]},
\label{GF-act}
\ee
up to surface terms, which do not concern us here.  Thus we see that in the dust time gauge,  the sum of the gravitational and matter parts of the Hamiltonian constraint  becomes the physical Hamiltonian,  and the diffeomorphism constraint reduces to that with only the gravity and matter $(\chi,p_\chi)$ contributions.  The ``vacuum'' theory, with $\chi=p_\chi=0$, has  six configuration degrees of freedom in $q_{ab}$, subject to the diffeomorphism constraint, giving three local  degrees of freedom. This is the action we study in the remainder of the paper. The corresponding spacetime metric is
\be
ds^2 = -dt^2 + (dx^a + N^a dt)(dx^b + N^b dt) q_{ab}. 
\ee 

\subsection{Deformation of the Hamiltonian}

So far we have described a matter time  gauge fixing of  canonical general relativity, which results in the action (\ref{GF-act}). Taking the latter as a starting point for defining theory, we  introduce a deformation of the gravitational Hamiltonian 
\be
 {\cal H}^G_\alpha := -\sqrt{q} R^{(3)}+\frac{1}{\sqrt{q}} \left(\tilde{\pi}^{ab}\tilde{\pi}_{ab}- \alpha \tilde{\pi}^{2}\right), 
\ee 
motivated by the Horava-Lifshitz models. In their original formulation, these models  are also constructed from a  first order action made from the spatial metric and extrinsic curvature; there is no covariant second order action as the starting point.   The models  also  have higher derivative 3-metric self-interactions through terms such as $R_{ab}R^{ab}$, as well as a deformation of the ADM kinetic term. The generalization we consider however only introduces  the latter through a parameter $\alpha$: 
 \be
S_\alpha=\frac{1}{2\pi}\int{ dt \ d^3x  \left[\tilde{\pi}^{ab}\dot{q}_{ab} +p_\chi\dot{\chi}  -( \mathcal{H}^{G}_\alpha + \mathcal{H}^{M}) -N^{a}(\mathcal{C}^{G}_{a} +\mathcal{C}^{M}_{a})  \right]}.
\label{Salpha}
\ee
In the following analysis we work with this deformation.  It has no effect on the constraint algebra since $\alpha$ is introduced only in the physical Hamiltonian.   We will see that for its GR value ($\alpha = 1/2$), the additional   degree of freedom in the metric  is ultralocal, whereas for  all other values it is a propagating scalar.  
 
\section{Linearized theory} 

For the remainder of this paper we assume that $\chi=p_\chi= \mathcal{H}^{M} = \mathcal{C}^M_a=0$, that is we consider the  action (\ref{GF-act}) without matter. It is easy to check that Minkowski spacetime, 
$q_{ab}=\delta_{ab}$, $\tilde{\pi}^{ab} =0 = N^a$,  is a solution of equations of motion in the dust time gauge. We  linearize the theory  on this background  by writing 
\be
q_{ab}(x,t) = \delta_{ab} + h_{ab}(x,t), \ \ \  \tilde{\pi}^{ab} = 0 + p^{ab}(x,t), \ \ \  N^a = 0+ \xi^{a}(x,t).
\ee

It is  convenient to work in $3-$momentum space by expanding the perturbations  $h_{ab}, p^{ab}, \xi^a$  in modes of the flat space Laplacian (plane waves) as
\be
h_{ab}(x,t)  &=& \frac{1}{(2\pi)^3}\int d^3k \  e^{ikx} \bar{h}_{ab}(k,t), \nn \\
p^{ab}(x,t) &=& \frac{1}{(2\pi)^3}\int d^3k \  e^{ikx} \bar{p}^{ab}(k,t)\nn\\
\xi^a(x,t) &=& \frac{1}{(2\pi)^3} \int d^3x\  e^{i kx} \bar{\xi}^a(k,t).
\ee
 This allows us to write the Hamiltonian and equations of motion   in Fourier space. The background solution $\delta_{ab}$ and  $k^a$  may be used to define an orthonormal basis of symmetric $3\times 3$ matrices $M^I$ so that the perturbations can be decomposed as 
\be
\bar{h}_{ab} = h_I(k,t) M^I_{ab}, \ \ \   \bar{p}^{ab} = p^I(k,t) M_I^{ab} ,\ \ \ \  I = 1,2 \cdots 6.
\ee
As we see below, the coefficients $(h_I, p^I)$ provide a natural separation of the perturbations into scalar, vector and tensor modes. Furthermore, if the chosen basis is static and orthonormal in the inner product 
\be
{\text Tr} (M^IM^J) = \delta^{IJ} \label{IP} ,
\ee
the symplectic form decomposes as   
\be
\int d^3k\  \bar{p}^{ab} \dot{\bar{h}}_{ab} = \int d^3k \  p^I \dot{h}_I. 
\ee
This identifies the six canonically conjugate degrees of freedom $ \displaystyle \left(h_I(k,t), p^I(k,t) \right)$.

A basis that fulfils these requirements is obtained by using an orthonormal basis of vectors 
\be
\label{bv}
\hat{k}^a = k^a/|k|,\ \ e_1^a, \ \  e_2^a,
\ee
where the latter pair span the plane orthogonal to $k^a$. By considering rotations $J_\alpha$ by angle $\alpha$ about the $k^a$-axis, one obtains a definition of `helicity' for the eigenvectors of these rotations; see eg. appendix A.2.1 in \cite{Baumann:2012}).  The eigenvectors of $J_\alpha$ are the linear combinations $e_{\pm}^a =  \left(e_1^a \pm ie_2^a\right)/\sqrt{2}$. These satisfy  $J_\alpha e^a_\pm = e^{\pm i\alpha} e^a_\pm$, $\delta_{ab} e^a_\pm e^b_\pm=0$ and $\delta_{ab}e^a_+ e^b_-=1$. The matrices 
\be
\label{bm}
\delta_{ab}, \ \ \  \hat{k}^a\hat{k}^b, \ \ \ e_\pm^{(a}\hat{k}^{b)}, \ \ \  e_\pm^a e^b_\pm.  \label{elements}
\ee
 are the eigentensors of $J_\alpha$: the first two are rotationally invariant and so are (helicity 0) scalars, the next pair are (helicity $\pm 1$) vectors, and the last pair are (helicity $\pm 2$) tensors.    

A basis $M^I$ with the above properties may be made as a linear combinations of these elements. We choose the scalar, tensor, and vector  bases respectively  as
\be
M_1^{ab} = \frac{1}{\sqrt{3}}\ \delta^{ab}, \ \ \ \ \  M_2^{ab} = \sqrt{\frac{3}{2}} \left(  \hat{k}^a\hat{k}^b -  \frac{1}{3} \delta^{ab}  \right),
\ee
\be
M_3^{ab} = \frac{i}{\sqrt{2}}\left( e^a_- e^b_-  - e^a_+e^b_+\right), \ \ \ \ \  M_4^{ab} = \frac{1}{\sqrt{2}}\left(e^a_- e^b_- + e^a_+e^b_+\right),
\ee
\be
M_5^{ab} =   i\left(e_-^{(a}\hat{k}^{b)} - e_+^{(a}\hat{k}^{b)} \right),  \ \ \ \ \  M_6^{ab} = \left(e_-^{(a}\hat{k}^{b)} +e_+^{(a}\hat{k}^{b)} \right).
\ee

The subset $M_I, I=2\cdots 6$ are trace free, $M^{ab}_I \delta_{ab}=0$,   and satisfy the traverse properties $k_a  M_3^{ab}=k_a M_4^{ab} =0$ and  $k_a k_bM_5^{ab} = k_a k_b M_6^{ab}=0$.  
(We note that the tensors of  definite helicity  in (\ref{elements}) have zero norm in the inner product (\ref{IP}) and lead to a degenerate reduction of the symplectic form. For this reason the above linear combinations of helicity tensors are necessary as basis elements in order to derive canonical equations of motion.)
  
Our goal now is to write the linearized canonical  Einstein equations in  the dust time gauge in $k-$space, fix three phase space  gauge conditions and solve the spatial diffeomorphism constraint. This will identify  the three local physical degrees of freedom. As we will see, two of these turn out to be the usual polarizations  of the graviton, and the third is the manifestation in the metric of the dust degree of freedom. The details of these steps follow.

\subsection{Linearized equations of motion}

The linearized equations about the flat background solution are
\be
\dot{h}_{ab}&=&2 \l(p_{ab} - \alpha \delta_{ab}p \r) + \mathcal{L_\xi}\delta_{ab} \nn \\
\dot{p}^{ab}&=& -\p^c\p^{(b} h^{a)}_{c}+\frac{1}{2}\p^c\p_ch^{ab} +\frac{1}{2}\p^a\p^b h+ \frac{1}{2}  \delta^{ab}\l(\p^c\p^dh_{cd}-\p^c\p_ch\r), 
\ee
 which in $k-$space are 
\be
\dot{\b{h}}_{ab}&=&2 \l(\b{p}_{ab} - \alpha\delta_{ab}\b{p} \r) + 2ik_{(a}\bar{\xi}_{b)}\nn \\
\dot{\b{p}}^{ab}&=&k^c k^{(b} \b{h}^{a)}_{c}-\frac{1}{2}k^c k_c\b{h}^{ab} -\frac{1}{2}k^a k^b \b{h}-\frac{1}{2}  \delta^{ab}\l(k^c k^d\b{h}_{cd}-k^c k_c\b{h}\r). 
\ee
 From these, the equations for the phase space pairs $(h_I,p^I)$ are obtained by projecting onto each basis element  $M^I$.  
The shift vector can be decomposed as
\be
\b{\xi}^a=\xi_{\parallel}\hat{k}^a+\xi_{1}e^{a}_{1}+\xi_{2}e^{a}_{2}. \label{shift}
\ee

 The scalar mode equations are 
\be
\dot{{h}}_{1}&=&2(1-3\alpha){p}_{1}+\frac{2i}{\sqrt{3}}|k|\xi_{\parallel},\\ 
\dot{{h}}_{2}&=&2{p}_{2}+2i\sqrt{\frac{2}{3}}|k| \xi_{\parallel},\\
\dot{{p}}_1 &=& \frac{1}{3} |k|^2 {h}_1 - \frac{1}{3\sqrt{2}} |k|^2 {h}_2 ,\\
\dot{{p}}_2 &=& -\frac{1}{3\sqrt{2}} |k|^2 {h}_1 + \frac{1}{6} |k|^2 {h}_2.
\ee
The tensor mode equations are 
\be
 \dot{{h}}_{3}&=&2{p}_{3}, \quad  \dot{{p}}_3 = -\frac{1}{2} |k|^2 {h}_3,  \label{TT1}\\
\dot{{h}}_{4}&=&2{p}_{4}, \quad  \dot{{p}}_4 = -\frac{1}{2} |k|^2 {h}_4,
\label{TT2}
\ee 
and the vector mode equations are 
\be
  \dot{{h}}_{5}&=&2{p}_{5}+i\sqrt{2}|k|\xi_{2},\quad \dot{{p}}_5 = 0,\\
\dot{{h}}_{6}&=&2{p}_{6}+i\sqrt{2}|k|\xi_{1}, \quad  \dot{{p}}_6 = 0.
\ee
These equations are supplemented by the linearized diffeomorphism constraint which we  discuss next. 
 
\subsection{Diffeomorphism constraint}

The position space diffeomorphism constraint $D_a\tilde\pi^{ab} =0$ linearizes about the flat background to $\p_a p^{ab}=0$. In $k$-space this is 
\be
 k_a \bar{p}^{ab} =  k_a p^I(k,t) M_I^{ab} = 0  \implies \left(  \frac{1}{\sqrt{3}} \ p_1  + \sqrt{\frac{2}{3}}\   p_2  \right) k^b+  \frac{|k|}{\sqrt{2}} \left( p_5\  e_2^b +  p_6\  e_1^b\right) =0. 
\ee 
It is evident that this constraint has transverse and longitudinal components, and furthermore,  that a partial  solution of this constraint must come from setting  $p_5=p_6=0$, since these are the only coefficients  in the transverse directions $e_1^a$ and $e_2^a$.

 More systematically, the vector modes are eliminated in three steps:  (i) imposing the gauge conditions  
\be
h_5=0, \ \ \ \  h_6=0, 
\ee
which are second class with the linearized diffeomorphism constraint, (ii)  solving the transverse component of the diffeomorphism constraint  by setting $p_5 = p_6 = 0$, and (iii)  using the conditions that the gauge be dynamically preserved to fix the transverse components of the shift perturbation (\ref{shift}),  
 \be
\dot{{h}}_{5}&=&i\sqrt{2}|k|\xi_{2}=0,\\
\dot{{h}}_{6}&=&i\sqrt{2}|k|\xi_{1}=0.
\ee
This fixes  $\xi_{1}=\xi_{2}=0$. The longitudinal component of the shift   $\b{\xi}^a=\xi_{\parallel}\hat{k}^a$ remains undetermined at this stage.

This leaves the scalar and tensor mode equations for  $(h_I,p_I), I=1\cdots 4$, and the longitudinal part of the  diffeomorphism constraint 
\be
\left(   \frac{1}{\sqrt{3}}p_1  + \sqrt{\frac{2}{3}}\   p_2  \right) =0. 
\ee
This remaining constraint is  on the two scalar degrees of freedom. After one more gauge fixing, the last of the three necessary to fully gauge fix the theory,  only one scalar mode and the 
transverse traceless graviton modes $(h_3,p_3)$ and $(h_4,p_4)$ remain. The former may be chosen as either the canonical pair $(h_1,p_1)$ or $(h_2,p_2)$.     

Let us consider the gauge $h_2=0$, and solve the remaining diffeomorphism constraint, giving $p_2 = -p_1/\sqrt{2}$.   The corresponding evolution equation gives  
\be
\dot{{h}}_{2}&=&2{p}_{2}+2|k|i\sqrt{\frac{2}{3}}\xi_{\parallel}=0 \implies \xi_{\parallel}= -i \frac{\sqrt{3}}{2|k|}\ p_1,
\ee
and the $\dot{p}_1$ and  $\dot{p}_2$ equations become identical. The remaining scalar mode equations reduce, using  the above expression for  $\xi_{\parallel}$, to 
\be
 \dot{{h}}_{1}&=& 3(1-2\alpha){p}_{1}  ,   \quad \quad \dot{{p}}_1 = \frac{1}{3} |k|^2 {h}_1,
\ee   
or equivalently,
\be
\ddot{{h}}_1 = (1-2\alpha) |k|^2 {h}_1. 
\ee
These are equivalent to the position space wave equation
\be
\ddot{h}_{1} =  (2\alpha-1)\  \delta^{ab} \p_a\p_b h_1,
\ee
so the propagation speed is $v = \sqrt{2\alpha -1}$. It is therefore evident that for the GR value $\alpha=1/2$, this  scalar mode is ultralocal:  there are no spatial derivatives in the equation, so  $h_1$ evolves independently  at  each space point. For   
$\alpha > 1/2$ the propagation speed varies from subluminal to superluminal, whereas for $\alpha<1/2$ the equation becomes a $4d$ Laplacian!  Had we chosen the gauge   ${h}_1=0$ (instead of $h_2=0$), a  similar analysis would reveal  the same wave equation for the scalar mode ${h}_2$.   

Lastly we note that the graviton (TT) modes (\ref{TT1})-(\ref{TT2}) are independent of $\alpha$ and satisfy the expected light speed wave equation
\be
 \ddot{{h}}_I  = - |k|^2 {h}_I,\ \ \  I = 3,4,
\ee
despite the dust time gauge fixing, which remarkably does not  effect Lorentz invariance in the linearized theory. This demonstrates that ``solving the problem of time" by adding a dust field is compatible with Lorentz invariance, and that the dust time gauge leads to no pathologies.

\subsection{Dust potential}
 
We note in passing that it is possible to  include a potential for the dust field in the starting theory \cite{Lim:2010yk}. This modifies the dust lagrangian to
\be
S^D = \int d^{4}x\sqrt{-g}\ \left[ m(g^{\mu\nu}\di_{\mu}\phi\di_{\nu}\phi+1)  - V(\phi)\right],
\ee 
and the dust contribution to the Hamiltonian constraint becomes
\be
\mathcal{H}^{D}=\frac{1}{2}\left(\frac{p_{\phi}^{2}}{m\sqrt{q}}+m\sqrt{q}(q^{ab}\di_{a}\phi\di_{b}\phi+1)\right) + \sqrt{q} \ V(\phi). 
\ee
Now the dust time gauge canonical action (\ref{Salpha}) becomes 
  \be
S_\alpha=\frac{1}{2\pi}\int{ dt \ d^3x  \left[\tilde{\pi}^{ab}\dot{q}_{ab} +p_\chi\dot{\chi}  -( \mathcal{H}^{G}_\alpha + \sqrt{q}\  V(t) + \mathcal{H}^{M}) -N^{a}(\mathcal{C}^{G}_{a} +\mathcal{C}^{M}_{a})  \right]}.
\ee
 This shows that the dust potential acts as a time dependent cosmological constant in the dust time gauge. It has been   studied  in explicit cosmological solutions in the context of mimetic gravity models \cite{Chamseddine:2013kea}.  
 
The consequences of  $V(t)$ for constructing a linearized theory are interesting. The first question is selecting a background solution on which to linearize the theory.  Minkowski space is no longer a solution due to the change in the equation for the ADM momentum $\tilde{\pi}^{ab}$.  Rather the simplest equations are cosmological for given $V(t)$, and the  analysis  differs significantly from  the flat  space linearized theory due to non-zero ADM momentum in the background solution. We leave this for future work, but note in particular that  the time dependent potential would drastically  affect the  graviton mode equation by introducing into it an explicit time dependence. This would obviously violate Lorentz covariance, which may be recoverable in epochs where $V(t)$ is chosen to be very slowing varying with $t$.  

\section{Discussion}
 
We studied general relativity coupled to pressureless dust in four spacetime dimensions, and analyzed in detail the linearized theory of perturbations about the flat spacetime solution in the ADM canonical theory.  This was done in the dust time gauge, which has the interesting feature that the physical Hamiltonian is particularly simple; it is the same function of phase space variables as the Hamiltonian constraint. We included  a one-parameter generalization of this canonical theory that resembles the Horava-Lifshitz model. 
 
 We find a number of interesting and surprising features: (i) the graviton modes satisfy a Lorentz invariant wave equation, despite the time gauge fixing, for any value of $\alpha$, (ii) the additional scalar mode in the metric is ultralocal for GR ($\alpha=1/2$), and so has no consequences associated with dynamical scalars, (iii) for   $\alpha>1/2$, the   scalar mode is propagating, but for $\alpha<1/2$, its equation becomes elliptic, (iv) inclusion of a dust potential provides a dynamical cosmological constant. 
 
 We note also that despite the ``intuition" from gravitational collapse of dust that the dust time gauge may break down due to shell crossing singularities, we have seen explicitly that  linearized theory in this gauge shows no hint of pathology.  Rather it shows that the ultralocality of the scalar mode (for the GR value $\alpha=1/2$) shields propagating modes, and may serve as a model for dark matter, for the simple reason that  ultralocal dynamics {\it is} dark! 
 
  In a similar study of gravity and dust in three spacetime dimensions \cite{Husain:2015RZ}, it was found that the scalar mode, the only local degree of freedom in 3D gravity, is ultralocal. It is interesting that the scalar perturbation is ultralocal in 4D as well, and leaves the usual graviton modes untouched. This decoupling of the scalar and tensor modes is a consequence of the special form of the gauge fixed action (\ref{GF-act}) which has the same spatial diffeomorphism constraint of GR coupled to matter fields. The structure of this constraint is such that at linear order the tensor modes remain unconstrained for any spatially constant metric. This happens also in the Einstein-Aether theories \cite{Jacobson:2004ts}. Moreover since the dust is a scalar, at linear order we expect the additional degree of freedom to be manifested only in the zero helicity   modes of the metric perturbations,  because of rotational invariance. 
 
Although we did not study matter fields other than the dust, at the linearized level their inclusion is straightforward. For example, if one includes a standard scalar field in the action, it is clear that perturbations about the zero solution would be Lorentz covariant. The same would hold for Fermi fields. The main point is that GR with dust, in the dust time gauge, appears to be perfectly consistent with standard Lorentz covariant field theory on Minkowski spacetime,   while at the same time ``solving the problem of time'' in quantum gravity. 

At the quantum gravity level we can write a time dependent functional Schrodinger equation 
\be
i\hbar \frac{\p \psi[q,\chi]}{\p t} = \int d^3x \left[    \hat{{\cal H}}^G  +    \hat{{\cal H}}^M       \right] \psi[q,\chi], \label{QG}
\ee 
a point discussed in the loop quantum gravity context in \cite{Husain:2011tk}, where a Hilbert space is available to precisely write down this equation; the cosmological setting is studied in  \cite{Husain:2011tm}. 

Combining these observations, we note in concluding that the dust time gauge provides compatibility between Lorentz invariance for linearized fields on Minkowski spacetime and  the  quantum gravity problem. But it remains to explore the quantum gravity equation (\ref{QG}) in settings larger than cosmology. Perhaps of  most interest is the problem of gravitational collapse, black hole formation, and  subsequent evolution in a fully quantum setting.

\begin{acknowledgments}

This work was supported by NSERC of Canada, and an AARMS Postdoctoral Fellowship to JZ. We thank Sanjeev Seahra for discussions.

\end{acknowledgments}

\bibliography{4dDust}

\end{document}